# Observation of Colossal Topological Hall Effect in Noncoplanar Ferromagnet $Cr_5Te_6$ Thin Films

Yequan Chen[1,2], Yingmei Zhu[3], Renju Lin[4], Wei Niu[1,5], Ruxin Liu[1], Wenzhuo Zhuang[1], Xu Zhang[1], Jinghua Liang[3], Wenxuan Sun[1], Zhongqiang Chen[1], Yongsheng Hu[2], Fengqi Song[2], Jian Zhou[6], Di Wu[2], Binghui Ge[4,*], Hongxin Yang[2,3,*], Rong Zhang[1,7], and Xuefeng Wang[1,*]

[1]Jiangsu Provincial Key Laboratory of Advanced Photonic and Electronic Materials, School of Electronic Science and Engineering, Collaborative Innovation Center of Advanced Microstructures, Nanjing University, Nanjing 210093, China
[2]National Laboratory of Solid State Microstructures, School of Physics, Collaborative Innovation Center of Advanced Microstructures, Nanjing University, Nanjing 210093, China
[3]Ningbo Institute of Materials Technology and Engineering, Chinese Academy of Sciences, Ningbo 315201, China
[4]Information Materials and Intelligent Sensing Laboratory of Anhui Province, Institutes of Physical Science and Information Technology, Anhui University, Hefei 230601, China
[5]School of Science, Nanjing University of Posts and Telecommunications, Nanjing 210023, China
[6]National Laboratory of Solid State Microstructures, Department of Materials Science and Engineering, Nanjing University, Nanjing 210093, China
[7]Department of Physics, Xiamen University, Xiamen 316005, China

*Authors to whom correspondence should be addressed. E-mail: bhge@ahu.edu.cn; hongxin.yang@nimte.ac.cn; xfwang@nju.edu.cn

**Abstract**
The topological Hall effect (THE) is critical to the exploration of the spin chirality generated by the real-space Berry curvature, which has attracted worldwide attention for its prospective applications in spintronic devices. However, the prominent THE remains elusive at room temperature, which severely restricts the practical integration of chiral spin textures. Here, we show a colossal intrinsic THE up to ~1.6 μΩ·cm in large-area ferromagnet $Cr_5Te_6$ thin films epitaxially grown by pulsed laser deposition. Such a THE can be maintained until 270 K, which is attributed to the field-stimulated noncoplanar spin textures induced by the interaction of the in-plane ferromagnet and antiferromagnet infrastructures. Our first-principles calculations further verify the considerable Dzyaloshinskii-Moriya interaction in $Cr_5Te_6$. This work not only paves the way for robust chiral spin textures near room temperature in large-area low-dimensional ferromagnetic films for practical applications, but also facilitates the development of high-density and dissipationless spintronic devices.

# 1. Introduction

Chiral spin textures generated by the real-space Berry curvature have attracted considerable interest in the past two decades owing to their promising spintronic applications.[1-9] They are often hosted in skyrmions, scalar spin chirality, noncollinear magnetic structures, and magnetic bubbles,[1-9] which are anticipated to be applied in next-generation room-temperature spintronic devices.[10] However, a comprehensive understanding of the exotic physics underlying chiral spin textures remains elusive. Specifically, the charge carriers are endowed with an extra phase factor owing to the emergent field created by chiral spin textures. An additional topological Hall effect (THE) usually emerges in the Hall resistance, which is distinct from the intrinsic anomalous Hall effect (AHE) caused by the Berry curvature in the momentum space.[9] Consequently, THE is considered an effective tool for the purely electrical detection of chiral spin textures.[9]

Since 2011, chiral spin textures have been extensively investigated through THE in various magnetic systems, including B20-type alloys (e.g., MnSi,[11] MnGe,[12] and FeGe[13] nanostripe), oxide films (e.g., $CrO_2$,[14] EuO,[15] $Ca_{1-x}Ce_xMnO_3$,[16] and $SrRuO_3$[17,18]), acentric tetragonal Heusler compounds (e.g., $Mn_{1.4}PtSn$[19] and $Mn_2RhSn$[20]), ferromagnetic heterostructures (e.g., $Tm_3Fe_5O_{12}$/Pt,[21] $LaMnO_3/SrIrO_3$,[22] and $BaFe_{12}O_{19}/Bi_2Se_3$[23]) and frustrated magnets (e.g., $Gd_2PdSi_3$[24]). However, simultaneously obtaining a colossal and high-temperature THE remains extremely challenging.

In recent years, there has been great interest in the burgeoning low-dimensional ferromagnetic materials,[25-27] such as the chromium telluride ($Cr_xTe_y$) family,[28-34] $Cr_2Ge_2Te_6$,[35] $Fe_3GeTe_2$,[36] $Fe_5GeTe_2$,[37] $Fe_3GaTe_2$,[38] and $CuCr_2Te_4$,[39] owing to their intrinsic high-temperature ferromagnetism for versatile applications.[40] The emergent THE has been notably found in these low-dimensional ferromagnetic systems, including their heterostructures (e.g., $CrTe/SrTiO_3$,[41] $Cr_2Te_3/Bi_2Te_3$,[42] $Fe_3GeTe_2/WTe_2$,[43] $CrTe_2/Bi_2Te_3$,[44] and $Cr_2Te_3/Cr_2Se_3$[45]), bulk forms (e.g., $Cr_5Te_8$,[46] $Cr_{0.9}B_{0.1}Te$,[47] $Cr_{0.87}Te$,[48] and $Cr_{1.53}Te_2$[49]), nanosheets (e.g., $Cr_5Si_3$[50]), and exfoliated nanoflakes (e.g., $Cr_{1.2}Te_2$[51]), demonstrating their potential application in chiral spintronics. However, the robust THE around room temperature in large-area homogeneous films is still insufficient considering the practical industrial applications based on chiral spin textures.[52]

In this work, we report a colossal intrinsic THE at high temperatures in large-area $Cr_5Te_6$ single-crystalline films grown by pulsed laser deposition (PLD). Its Curie temperature ($T_C$) is as large as 320 K. The THE resistivity is observed up to 1.6 µΩ·cm at 90 K, which is the largest magnitude in the $Cr_xTe_y$ family. Remarkably, this THE magnitude can be maintained until 270 K, suggesting the robust chiral spin textures existing in our $Cr_5Te_6$ films. The interaction of the in-plane ferromagnet and antiferromagnet infrastructures generates critical noncoplanar spin textures under the out-of-plane magnetic field, leading to the colossal THE. The considerable Dzyaloshinskii–Moriya interaction (DMI) in $Cr_5Te_6$ is verified by our first-principles calculations. These findings facilitate the development of room-temperature chiral spintronic devices based on large-area $Cr_5Te_6$ thin films.

## 2. Results and Discussion

The experimental setup for the epitaxial growth of the $Cr_5Te_6$ films (5 × 5 mm$^2$) by PLD is illustrated in Figure 1a. The $Cr_5Te_6$ possesses a monoclinic structure with the space group of $I2/m$-$C_{2h}^3$, very similar to that of CrTe.[53,54] In contrast, there are two types of Cr atoms in $Cr_5Te_6$ (Figure 1a), $Cr_1$ and $Cr_2$, represented by the blue and green solid spheres, respectively. The layers of $Cr_1$ and $Cr_2$ are stacked alternately along the $c$ axis.[55] Figure 1a also shows the reflection high-energy electron diffraction (RHEED) patterns of the $Cr_5Te_6$ films on $Al_2O_3$ (001) substrates after growth. The distinct three-dimensional (3D) dotted matrix patterns suggest that the crystallization of $Cr_5Te_6$ films is 3D island-like growth. The Raman spectrum of the $Cr_5Te_6$ films reveals two main phonon peaks at approximately 125 cm$^{-1}$ and 142 cm$^{-1}$ (see Figure S1a, Supporting Information), corresponding to $A_{1g}$ and $E_g$ vibrational modes of the $Cr_xTe_y$ family, respectively.[56] In addition to the $Al_2O_3$ substrate, the x-ray diffraction (XRD) patterns exhibit diffraction peaks of the (002$n$) family of planes (see Figure S1b, Supporting Information), indicating the surface perpendicular to the $c$ axis of $Cr_5Te_6$.[55] The atomic force microscope (AFM) image is also provided (Figure S1c, Supporting Information), indicating the relatively flat surface of the $Cr_5Te_6$ films. To evaluate the crystal structure, cross-sectional high-angle annular dark field-scanning transmission electron microscopy (HAADF-STEM) is performed on the $Cr_5Te_6$ (20 nm) film with the two-dimensional (2D) integrated differential phase contrast (iDPC) technique. The energy dispersive spectroscopy (EDS) maps in Figure 1b show the elemental distribution of Al, Cr, and Te in the sample, suggesting the negligible elemental interdiffusion at the interface of $Cr_5Te_6/Al_2O_3$. The stoichiometric ratio of Cr:Te is 45.44:54.56, which coincides with that of $Cr_5Te_6$. Figures 1c and 1d illustrate the HAADF-STEM and iDPC images near the interface viewed along the [110] direction, respectively. The well-stacked crystal lattice demonstrates the high-quality epitaxy of the $Cr_5Te_6$ films on sapphire. The iDPC image (Figure 1d) exactly matches the crystal structures of $Cr_5Te_6$. Moreover, the selected-area electron diffraction (SAED) pattern along the [110] zone axis further nails down the single-crystalline nature of the $Cr_5Te_6$ films (Figure 1e).

The magnetic properties of the $Cr_5Te_6$ films (both in-plane and out-of-plane configurations) were measured using superconducting quantum interference device magnetometer (SQUID). The curves in Figure 2a show the temperature dependence of the magnetization ($M$-$T$) of the 20-nm-thick $Cr_5Te_6$ film under an in-plane and out-of-plane magnetic field of 1000 Oe, respectively. The out-of-plane $M$-$T$ curve shows a general decreasing trend with increasing temperature, whereas there is a notable bump at 112 K in the in-plane $M$-$T$ curve. This indicates that in addition to the dominant ferromagnetic phase, a potential antiferromagnetic phase parallel to the film surface exists. Analogous bumps in the $M$-$T$ curves were also observed, and they were assigned to the Cr vacancies in the bulk $Cr_5Te_6$.[55] The peak of the bump separates the regions $P_1$ (mixed ferromagnetic and antiferromagnetic phase) and $P_2$ (ferromagnetic phase) in the $M$-$T$ curve. Based on the scaling law $M \sim (T_C-T)^\beta$,[28] the data near $T_C$ are fitted, where $T_C$ and the critical exponent $\beta$ are calculated to be 320 K and 0.476, respectively. Compared with films of similar stoichiometries, e.g., $Cr_4Te_5$ ($T_C$~260 K)[57] and CrTe ($T_C$~200 K)[41] films, the $Cr_5Te_6$ films have the highest $T_C$ above room temperature. The in-plane and out-of-plane magnetic hysteresis loops of the 20-nm-thick $Cr_5Te_6$ film at different temperatures demonstrate that the magnetic easy axis is along the in-plane direction (see Figures S2, Supporting Information). The average atomic magnetic moment is 0.45 $\mu_B$/Cr at 300 K, which is a typically high value in the $Cr_xTe_y$ family owing to the large Cr/Te compositional ratio of $Cr_5Te_6$.

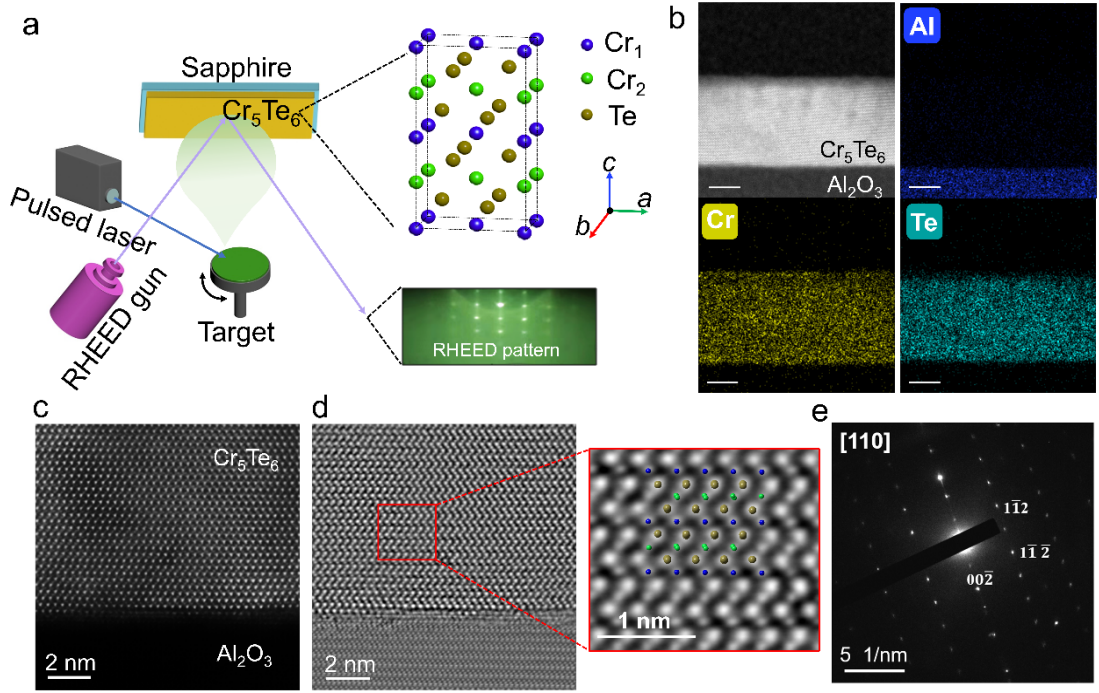

**Figure 1.** Large-area fabrication of $Cr_5Te_6$ films and microscopic characterization. a) Schematic diagram of the PLD setup, the crystal structure of $Cr_5Te_6$ and the RHEED pattern of $Cr_5Te_6$ films after growth. The dashed box marks the unit cell of $Cr_5Te_6$. b) The EDS elemental analysis of the cross-sectional specimen (including Al, Cr and Te). The scale bar is 10 nm. c,d) Atomic-resolution HAADF-STEM image and the iDPC image taken along [110] direction, respectively. The right panel shows the corresponding crystal structures with $Cr_1$, $Cr_2$ and Te atoms in blue, green and grey, respectively. e) The SAED pattern taken along the [110] zone axis of the $Cr_5Te_6$ films.

To explore the magnetotransport properties, the $Cr_5Te_6$ films with thicknesses of 20 and 50 nm are patterned into the standard Hall-bar geometry (Figure 2b). The temperature-dependent longitudinal resistances $R_{xx}$ show that 50-nm-thick $Cr_5Te_6$ exhibits the metallic behavior, and is thus selected for further low-temperature transport measurements (Figure S3, Supporting Information). The hysteresis loop of Hall resistivity ($\rho_{xy}$) displays a remarkable square shape at low temperatures, whereas the coercive field ($H_C$) decreases with increasing temperature (Figure 2c). Moreover, the distinct remaining $\rho_{xy}$ at 300 K implies a $T_C$ above room temperature. The red and blue curves represent $\rho_{xy}$ with upward and downward field sweeps, respectively. Notably, an anomalous hump attached to the hysteresis loop is clearly observed over a broad temperature range (10-250 K), which is a sharp antisymmetric peak around $H_C$ followed by a smooth suppression at the high field. Typically, the hysteresis loop of

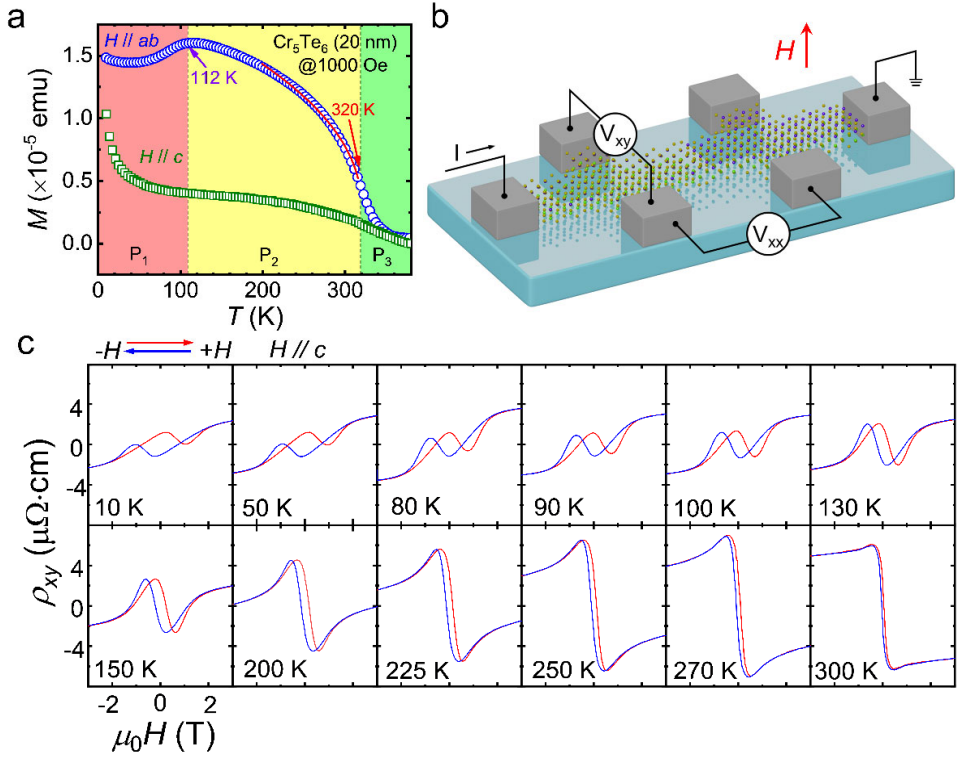

**Figure 2.** Magnetic and magnetotransport properties of $Cr_5Te_6$ thin films. a) The temperature-dependent magnetization curves of the 20-nm-thick $Cr_5Te_6$ films with applied field parallel to the $c$ axis and $ab$ plane, respectively. Three regions are denoted in the $M$-$T$ curve: $P_1$ (mixed ferromagnetic and antiferromagnetic phase), $P_2$ (ferromagnetic phase), and $P_3$ (paramagnetic phase). b) Schematic illustration of the Hall bar device structure with the six-terminal geometry. c) The field-dependent Hall resistivity of the 50-nm-thick $Cr_5Te_6$ films at various temperatures.

$\rho_{xy}$ in ferromagnetic materials is composed of the ordinary Hall effect (OHE) and the AHE. The OHE is due to the Lorentz force acting on the charge carriers, whereas the AHE is due to the magnetization. The distinct hump indicates the extra THE in the $Cr_5Te_6$ films produced by the topological spin chirality structure. This complex physics is achieved when charge carriers travel through the emergent magnetic field, which is a fictitious field generated by the Berry phase effect in real-space.[16,21] However, several previous studies held a different viewpoint that the superposition of two AHE hysteresis loops with the different $H_C$ may result in a similar anomalous hump (i.e., two-AHE model).[58-61] Two cases of the superposition are provided (see Figures S4a and S4c, Supporting Information), that is, the dominant (minor) AHE with low (high) $H_{C1}$ ($H_{C2}$) and the dominant (minor) AHE with high (low) $H_{C2}$ ($H_{C1}$), respectively. It can be observed that the anomalous hump is located between $H_{C1}$ and $H_{C2}$. This artificial signal is mainly attributed to two magnetic domains (magnetic phase) caused by inhomogeneity and/or heterostructure rather than the Berry phase. Therefore, it is vital to distinguish whether the hump originates from a genuine THE or the two-AHE model.

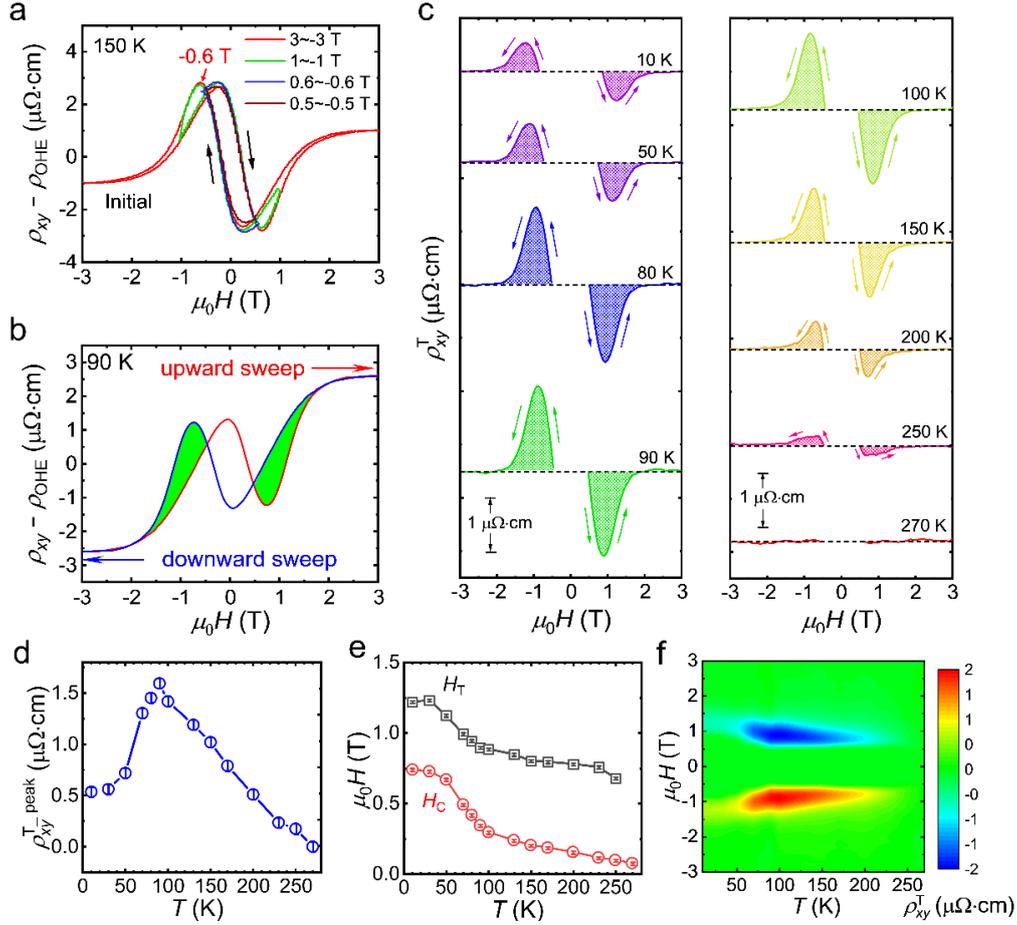

**Figure 3.** The THE in $Cr_5Te_6$ thin films with the thickness of 50 nm. a) The Hall measurement of minor loops ($\rho_{xy} - \rho_{OHE}$) with the initial magnetic field of -3 T at 150 K. b) The magnetic-field dependent Hall resistivity ($\rho_{xy} - \rho_{OHE}$) at 90 K, in which the red and blue curves represent the data of upward sweep and downward sweep, respectively. The green area is the difference between the red and blue curves, marking the extracted THE component. c) The typical pairs of field dependence of the topological Hall resistivity at various temperatures (10-270 K). d) The temperature dependence of the magnitude of $\rho_{THE}$. e) The temperature dependence of the field at which the topological Hall resistivity reaches its maximum ($H_T$) and the coercive field ($H_C$) of the AHE, respectively. f) The experimental color map of the topological Hall resistivity in the temperature-magnetic field plane.

In our case, the Hall measurement of minor loops is implemented on a 50-nm-thick $Cr_5Te_6$ film, which can justifiably demonstrate the genuine THE.[62,63] Before the measurements of each minor loop, the external field is set at −3 T to ensure the saturation magnetization of the $Cr_5Te_6$ films. Subsequently, the external field sweeps a minor loop with a field smaller than 3 T. Given the position of the hump (~ 0.6 T), we choose 1.0 T, 0.6 T, and 0.5 T as the field of the minor loops (Figure 3a). Obviously, the positions of all three minor Hall loops are unchanged, along with the similar $H_C$ values. Based on the two-AHE model, two opposite magnetic domains must be present in the sample, defined as $M_1$ with the small $H_{C1}$ and $M_2$ with the large $H_{C2}$. After the initial high-field magnetization, which is marked as "Initial" (see Figures S4b and S4d, Supporting Information), both $M_1$ and $M_2$ reach their saturation states. Subsequently, the external magnetic field gradually sweeps a minor loop with the field between $H_{C1}$ and $H_{C2}$. During this process, $M_1$ reverses with the field sweep but $M_2$ is pinned. Hence, the superposition of two AHE hysteresis loops is equal to the sum of the AHE hysteresis loop of $M_1$ and a constant value, as described by the black loops (see Figures S4b and S4d, Supporting Information). Compared with the cases shown in Figures S4a and S4c, the minor Hall loops demonstrate a different $H_C$ and an extra offset. Consequently, the unchanged position and $H_C$ of the minor Hall loops (Figure 3a) show the genuine THE in the $Cr_5Te_6$ films. Additionally, we use a double tanh function to fit the possible two-AHE hysteresis loops in the Hall measurement (see Figure S5, Supporting Information),[44] further ruling out the two-AHE model.[44]

In addition to the OHE and AHE, THE resulting from the real-space Berry phase contributes to the $\rho_{xy}$ of the $Cr_5Te_6$ films, which can be expressed as follows:

$$\rho_{xy}(H) = \rho_{OHE} + \rho_{AHE} + \rho_{xy}^{T} \tag{1}$$

where $\rho_{OHE}$ is the OHE part, $\rho_{AHE}$ is the AHE component, and $\rho_{xy}^{T}$ represents the contribution of the THE. Two methods are adopted to extract the THE component: (1) using a linear fitting and a step function, $M_0 \tanh\left(\dfrac{H}{a_0} - H_0\right)$, to subtract the OHE and AHE components;[18] and (2) taking the difference of $\rho_{xy}$ between the upward and downward field sweeps.[64] The clear difference between the upward and downward field sweeps is observed around the humps in all hysteresis loops of $\rho_{xy}$ below 270 K (Figure 2c). Thus, we directly obtain THE component by taking the difference of $\rho_{xy}$ between the upward and downward field sweeps (Figure 3b). Figure 3c shows the field dependence of THE component at different temperatures, which exhibits a hysteresis behavior. To further explore the trend of the THE hump, the peaked value of the hump, $\rho_{xy}^{T\_peak}$, the position of the peaked value $H_T$ (defined in Figure S6, Supporting Information), and $H_C$ as a function of temperature are plotted in Figures 3d and 3e, respectively, where the error bars are determined by the resolution of the measurement system. At low temperatures, $\rho_{xy}^{T\_peak}$ continues to increase with increasing temperature and reaches the maximum of ~1.6 μΩ·cm at 90 K (very close to the phase transition point between $P_1$ and $P_2$), which is the largest value among the reported $Cr_xTe_y$ family.[41,42,44-49,51] This suggests a strong coupling between charge carriers and the topological spin textures in the $Cr_5Te_6$ films around the phase transition point.[65] Subsequently, $\rho_{xy}^{T\_peak}$ gradually decays with increasing temperature, which is sizable till 270 K. Above 270 K, the sole AHE is observed, indicating the disappearance of spin chirality at higher temperatures. A similar nonmonotonic change in the THE peaks has been reported in many systems, e.g., $CrTe/SrTiO_3$,[41] $Cr_5Te_8$,[46] $SrRuO_3/SrTiO_3$,[66] and $EuIn_2As_2$.[65] The curves of temperature-dependent $H_T$ and $H_C$ show similar trends (Figure 3e), suggesting that spin chirality is induced when the magnetic moments are reversed. Figure 3f plots a phase diagram of THE curves of all temperatures and fields. The colossal THE at 90 K (~1.6 μΩ·cm) indicates the strong fictitious magnetic field induced by the Berry phase in real space. Moreover, the robust topological spin textures in the $Cr_5Te_6$ films are manifested by the high-temperature THE hump.

**Table 1.** Comparison of the maximal topological Hall resistivity and the highest temperature of the observed THE in our $Cr_5Te_6$ thin films and other systems.

| Material system | | Maximum THE (μΩ·cm) | Maximum temperature (K) | Spin structure | References |
|---|---|---|---|---|---|
| | $Cr_5Te_6$ film | 1.6 | 270 | Noncoplanar spin textures | This work |
| $Cr_xTe_y$ system | $CrTe_2/Bi_2Te_3$ | 1.39 | 120 | Interfacial DMI | [44] |
| | $CrTe/SrTiO_3$ | 1.2 | 75 | Néel-type skyrmions | [41] |
| | $Cr_2Te_3/Bi_2Te_3$ | 0.5 | 40 | Interface skyrmions | [42] |
| | $Cr_2Te_3/Cr_2Se_3$ | 0.23 | 75 | Noncoplanar spin textures | [45] |
| | $(Cr_{0.9}B_{0.1})Te$ | 0.215 | 140 | Néel-type skyrmions | [47] |
| | $Cr_{0.87}Te$ | 0.2 | 298 | Skyrmionic Bubbles | [48] |
| | $Cr_{1.53}Te_2$ | 0.106 | 290 | Skyrmions | [49] |
| | $Cr_{1.2}Te_2$ | 0.05 | 300 | Possible THE | [51] |
| | $Cr_5Te_8$ | 0.016 | 160 | Noncoplanar spin textures | [46] |
| Other typical systems | $Ca_{0.99}Ce_{0.01}MnO_3$ | 120 | 75 | Bubble-type skyrmions | [16] |
| | $LaMnO_3/SrIrO_3$ | 75 | 50 | Interface skyrmions | [22] |
| | EuO | 12.4 | 70 | Two-dimensional skyrmions | [15] |
| | $Gd_2PdSi_3$ | 2.6 | 20 | Hexagonal skyrmions | [24] |
| | $SrRuO_3$ | 0.4 | 80 | Interfacial DMI | [17] |
| | MnGe | 0.3 | 200 | Hexagonal skyrmions | [12] |
| | $BaFe_{12}O_{19}/Bi_2Se_3$ | 0.24 | 100 | Interfacial DMI | [23] |
| | MnSi | 0.01 | 45 | Hexagonal skyrmions | [11] |
| | $Tm_3Fe_5O_{12}/Pt$ | 0.005 | 400 | Interfacial DMI | [21] |

The maximal magnitudes as well as the highest temperatures of the observed THE of different materials and heterostructures are summarized in Table 1, including B20-type magnetic materials,[11,12] oxide films,[15-17] magnetic heterostructures,[21-23] frustrated magnets,[24] and the $Cr_xTe_y$ family.[41,42,44-49,51] The maximal value of THE in our $Cr_5Te_6$ films (~1.6 μΩ·cm) is extraordinarily high compared with those of most typical materials and structures. More importantly, this THE can be maintained up to near room temperature (270 K), indicating the potential for future applications in chiral spintronic devices. The temperature-dependent THE magnitudes of several typical material systems further display the enormous potential of our $Cr_5Te_6$ films in chiral spintronics, as summarized in Figure 4.

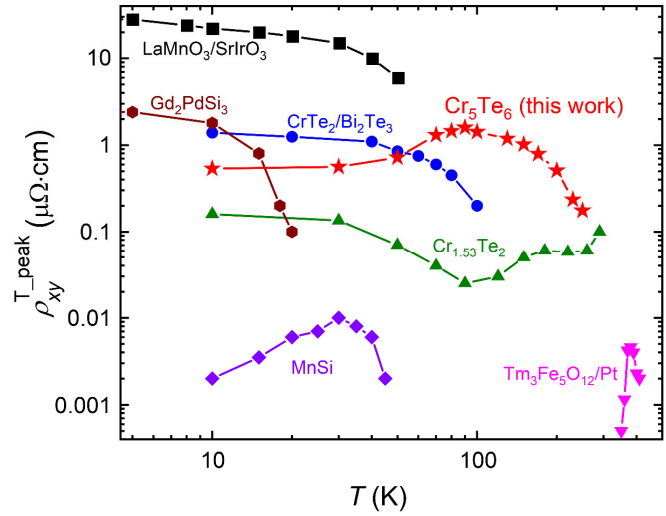

**Figure 4.** Summary of the topological Hall resistivity in the different systems. The temperature dependences of the observed THE magnitudes in $Cr_5Te_6$ films (this work) and other systems, such as MnSi,[11] $Tm_3Fe_5O_{12}$/Pt,[21] $LaMnO_3$/$SrIrO_3$,[22] $Gd_2PdSi_3$,[24] $CrTe_2$/$Bi_2Te_3$,[44] and $Cr_{1.53}Te_2$.[49]

In general, there are several mechanisms responsible for the topological spin textures resulting in the THE hump in Hall measurements, such as the bulk DMI in non-centrosymmetric structures,[11,12] interfacial DMI in heterostructures of ferro-/ferri-/antiferro-magnetic layers and heavy-metal layers,[21,67] geometrical frustration interaction in frustrated magnets,[68,69] and competition of various magnetic interactions caused by composition, thickness, electric field, and thermal effect.[9] Notably, few stoichiometries among the $Cr_xTe_y$ family, e.g., the $Cr_5Te_6$ films in our work and bulk $Cr_{0.87}Te$,[48] exhibit the intrinsic out-of-plane THE. Thus, the origin of the colossal THE in the $Cr_5Te_6$ films is addressed. According to the neutron diffraction results of $Cr_5Te_6$, two different types of Cr atoms possess four different orientations of spin vectors owing to the interaction of the in-plane ferromagnetism and antiferromagnetism infrastructures (Figure 5a).[54,70] The in-plane ferromagnetic component is along the *a* axis, whereas the antiferromagnetic component is arranged along the *b* axis. All the spins lie in the *ab* plane without the external magnetic field. When the applied field is less than $H_C$, the spin is oriented with an extra *c*-axis component, inducing the noncoplanar spins with scalar spin chirality in the $Cr_5Te_6$ films. This acts as a fictitious magnetic field along the *c* axis and leads to the observed THE in the Hall measurements. With an external field larger than $H_C$, all the spins are aligned into the *c* axis, resulting in the vanishing of noncoplanar spin textures and the THE. Notably, an in-plane THE (in $\rho_{zx}$) was reported previously in $Cr_5Te_8$ owing to the similar field-induced noncoplanar spin textures.[46]

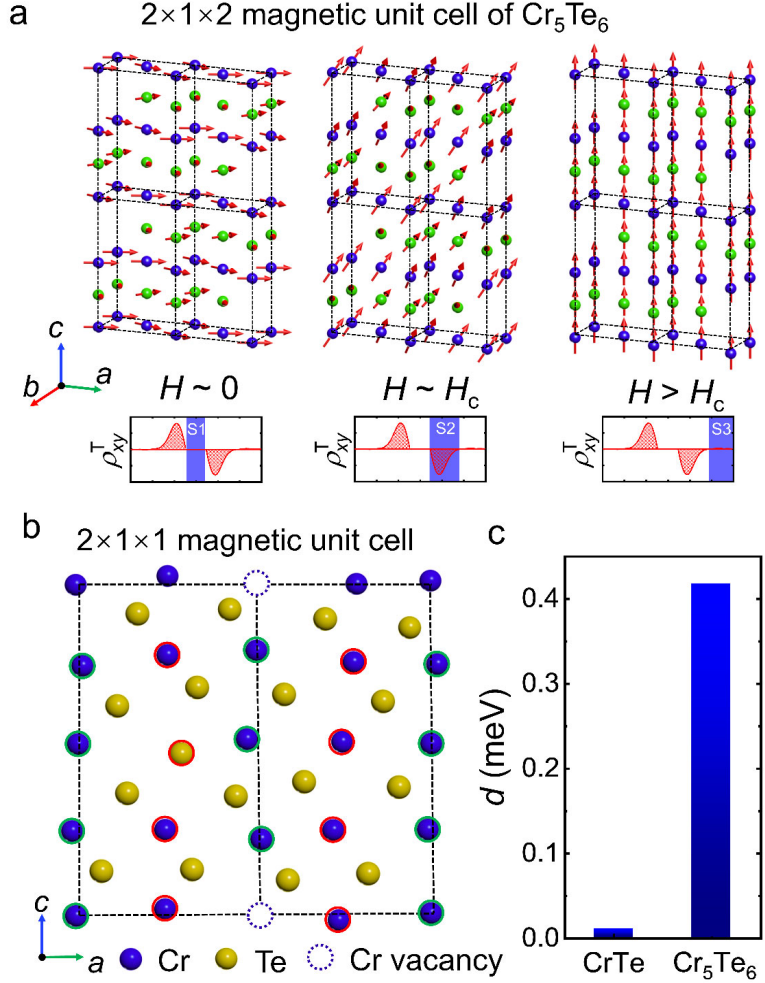

**Figure 5.** The THE induced by the noncoplanar spin textures. a) Spin configuration of the Cr for 2×1×2 supercell with an out-of-plane magnetic field of $H\sim0$, $H\sim H_c$ and $H>H_c$, respectively. The corresponding regions in the $\rho_{xy}^T$ versus magnetic field are marked by the blue rectangles, named as S1, S2 and S3, respectively. b) The constructed 2×1×1 supercell of $Cr_5Te_6$ with 1.2% vacancy of Cr. The red (green) circles stand for the atoms possessing the opposite (same) spin orientations with ACW and CW chiral spin configurations, respectively. c) The calculated DMI strength (*d*) of $Cr_5Te_6$ and CrTe.

To fully clarify THE in the $Cr_5Te_6$, the DMI in the $Cr_5Te_6$ system is quantified by the first-principles calculations. Owing to the potential Cr vacancy in the $Cr_xTe_y$ system and the stoichiometric ratio of Cr:Te determined in $Cr_5Te_6$ films (Figure 1b), we construct a 2×1×1 supercell of $Cr_5Te_6$ with 1.2% vacancy of Cr element based on the lattice structure of CrTe (Figure 5b). All the atoms, except for the four at the top, are considered during the calculations. The red (green) circles denote the atoms holding the opposite (same) spin orientations with anticlockwise (ACW) and clockwise (CW) chiral spin configurations (see Figure S7, Supporting Information). The DMI strength ($d$) is calculated by comparing the energy difference between ACW and CW chiral spin configurations using the following equation:[71]

$$d = (E_{ACW} - E_{CW})/8\sqrt{3} \qquad (2)$$

The extracted $d$ of $Cr_5Te_6$ is 0.42 meV, whereas that of CrTe is close to 0.00 meV (Figure 5c), indicating that the former system favors the critical noncollinear magnetic configuration, and thus, produces the robust THE. Furthermore, the HE in the $Cr_5Te_6$ films is very different from that in the 2D skyrmions (see Figure S8, Supporting Information).[15] The spin chirality caused by thermal fluctuation is also ruled out according to the disappeared THE above 270 K, which is below the $T_C$ of $Cr_5Te_6$.[18] Therefore, both the experimental evidence and theoretical calculations corroborate the existence of the THE in the $Cr_5Te_6$ films, which can be attributed to the noncoplanar spin textures induced by the interaction of the in-plane ferromagnetism and antiferromagnetism infrastructures.

## 3. Conclusion

In summary, we have fabricated large-area $Cr_5Te_6$ single-crystalline films by the PLD. A colossal intrinsic THE (~1.6 μΩ·cm at 90 K) is clearly observed, which is the largest magnitude among the $Cr_xTe_y$ family. Notably, the survival of THE near room temperature (270 K) demonstrates the robust chiral spin textures in the $Cr_5Te_6$ films. The THE is driven by the noncoplanar spin textures originating from the interaction of the in-plane ferromagnet and antiferromagnet infrastructures. Furthermore, the first-principles calculations well explain the underlying mechanism, suggesting the higher survival temperatures of the THE upon the future film-quality optimization. Our PLD methodology offers a promising platform of large-area $Cr_5Te_6$-based heterostructures for the development of chiral spintronic devices. The findings pave the pathway for the manipulation of chiral spin textures at the higher temperatures and accelerate the practical applications based on chiral spintronics.

## 4. Experimental Section

**Large-Area Epitaxy of $Cr_5Te_6$ Thin Films**: The target was prepared by heating mixed chromium (99.99%) and tellurium (99.99%) powders with a stoichiometric ratio of 1:3 at 450 °C for four days. The basic pressure of the PLD vacuum chamber was approximately 2.9 × 10$^{-5}$ Pa, and the distance between the substrate and target was 5 cm. The sapphire substrates (5 × 5 mm$^2$) were cleaned in turn by ultrasonic running acetone, alcohol, and deionized water for 5 min, in that order. The $Cr_5Te_6$ films were deposited onto the substrates at 500 °C at a speed of 1 nm/min using a 248-nm KrF excimer laser beam (an average fluence of 1 J/cm$^2$ and a repetition rate of 2 Hz). The growth process was monitored by RHEED patterns to ensure the high quality of thin films.

**Structural Characterizations**: The crystalline structure of the $Cr_5Te_6$ films was determined by a micro-Raman spectrometer (NT-MDT nanofinder-30) with a 514.5 nm wavelength $Ar^+$ laser as an excitation source in the backscattering configuration. The crystallography was examined by $\theta$-$2\theta$ XRD (Rigaku D/MAX-Ultima III) using Cu $K_\alpha$ radiation. The surface morphology was swept by an AFM system (NT-MDT). The sample for cross-sectional STEM observation was prepared on a Carl Zeiss crossbeam 550L FIB+SEM dual beam system using the conventional lift-out method. An electron diffraction pattern was accomplished using a JEOL-F200. The HAADF, iDPC-STEM imaging, and EDS mapping were performed using a FEI Titan Themis Z microscope equipped with probe and image correctors operated at 300 kV. The HAADF-STEM images were acquired using the convergence angle of 25 mrad and the collection angle of 50-200 mrad.

**Magnetic Measurements**: Magnetic properties were measured by a Quantum Design SQUID system. The temperature-dependent magnetization curves were measured from 10 to 300 K with the magnetic field (1000 Oe) applied in-plane and out-of-plane of the substrates, respectively.

**Magnetotransport Measurements**: Magnetotransport properties were measured in the standard Hall-bar geometry ($2.5 \times 0.5$ mm$^2$) with ultrasonically wire-bonded aluminum wires as electrodes. A Cryogen Free Measurement System (CFMS, Cryogenic) was employed with temperatures ranging from 5 to 300 K in a perpendicular magnetic field up to 3 T. The Hall resistivity $\rho_{xy}$ was calculated from the raw data as follows:

$$\rho_{xy} = -t\left[V_H(+H \to -H) - V_H(-H \to +H)\right]/2I \qquad (3)$$

where $t$ is the film thickness and $I$ is the excitation current. The offsets for Hall resistivity can be effectively subtracted owing to misalignment of contacts.[66]

**First-Principles Calculations**: The first-principles calculations were performed by the density functional theory (DFT) using the Vienna ab initio Simulation Package (VASP) code.[72,73] The exchange and correlation functionals were implemented by the generalized gradient approximation (GGA) of the Perdew-Burke-Ernzerhof functional.[74] Structures were fully relaxed until the force converged on each atom less than $10^{-2}$ eV/Å, and the convergence precision of total energy was set to $10^{-7}$ eV. The plane-wave cutoff energy was set to 350 eV. The Brillouin zone was sampled using a Γ-centered $5 \times 20 \times 5$ Monkhorst-pack $k$ mesh[75] in our calculations of DMI for a $2 \times 1 \times 1$ supercell (see Figure S7, Supporting Information). The real space spin–spiral energy difference approach was adopted to calculate the DMI strength.[71,76]

**Acknowledgements**
This work was supported in part by the National Key R&D Program of China (grant nos. 2022YFA1402404 and 2022YFA1405102), the National Natural Science Foundation of China (grant nos. 62274085, 11874203, 61822403, and 12174405), the

Fundamental Research Funds for the Central Universities (grant no. 021014380080), and the Micro-Fabrication and Integration Technology Center of Nanjing University.

**Supporting Information**

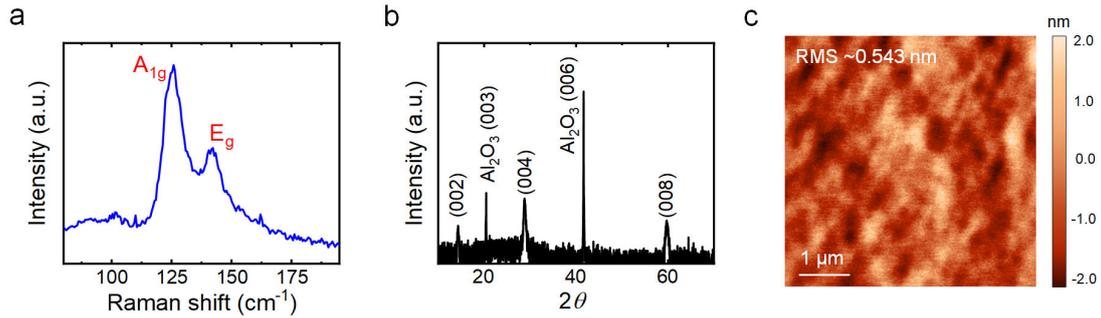

**Figure S1.** a) The Raman spectrum of $Cr_5Te_6$ films. b) The typical XRD pattern of $Cr_5Te_6$ films. The Raman spectrum of the $Cr_5Te_6$ films reveals two main phonon peaks at about 125 cm$^{-1}$ and 142 cm$^{-1}$, corresponding to $A_{1g}$ and $E_g$ vibrational modes of the $Cr_xTe_y$ family.[1] Besides the $Al_2O_3$ substrate, the XRD pattern exhibits diffraction peaks of (002$n$) family of planes.[2] c) The AFM image of the 20-nm-thick $Cr_5Te_6$ films with the roughness of ~0.543 nm. RMS means root mean square.

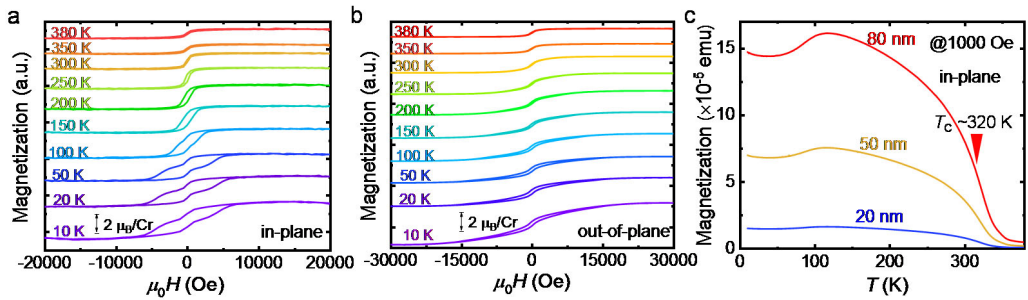

**Figure S2.** a,b) The magnetic hysteresis loops of the 20-nm-thick $Cr_5Te_6$ film with the in-plane and out-of-plane field at different temperatures, respectively. It demonstrates that the magnetic easy axis is along the in-plane direction. c) The $M$-$T$ curves of the $Cr_5Te_6$ films with the in-plane field (1000 Oe). For the films with thickness of 20, 50, and 80 nm, the $T_C$ of ~320 K remains unchanged.

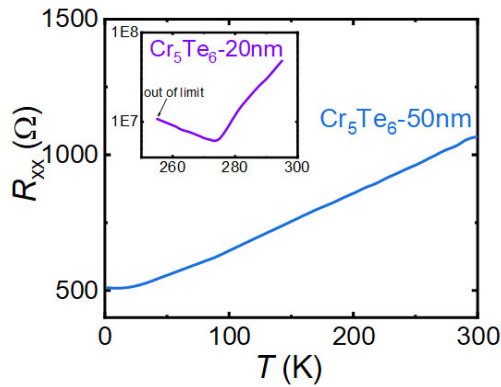

**Figure S3.** The temperature-dependent longitudinal resistance $R_{xx}$ of the $Cr_5Te_6$ films with the different thickness. The 20-nm-thick $Cr_5Te_6$ (inset) shows the insulating behavior while the 50-nm-thick $Cr_5Te_6$ exhibits the metallic behavior.

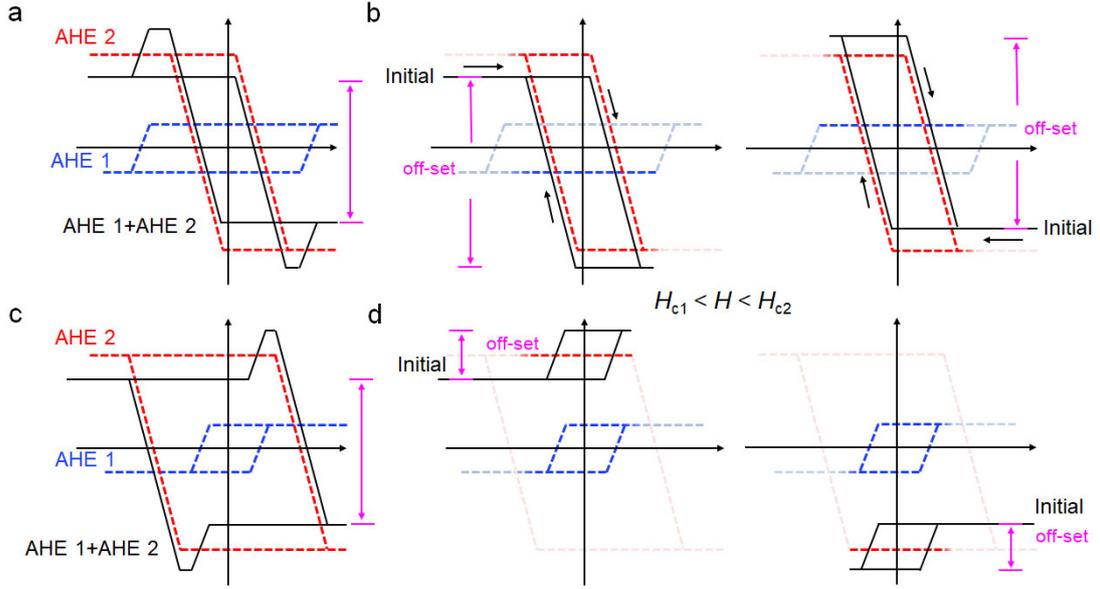

**Figure S4.** a,c) Two cases of the superposition based on the two AHE model. b,d) The results of minor loops magnetic field Hall measurement based on the cases in (a) and (c), respectively. Compared with the cases in (a) and (c), the minor Hall loops in (b) and (d) show the different $H_C$ and an extra off-set, respectively, very different from the results in Figure 2d of the main text.

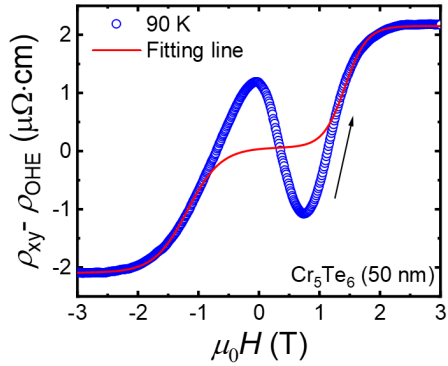

**Figure S5.** Fitting result of the field-dependent Hall resistivity ($\rho_{xy} - \rho_{OHE}$) at 90 K using the double tanh function. The fitting result, especially the data around the position of the hump in the $\rho_{xy}$, indicates that the two AHE model cannot fit the $Cr_5Te_6$ film well.

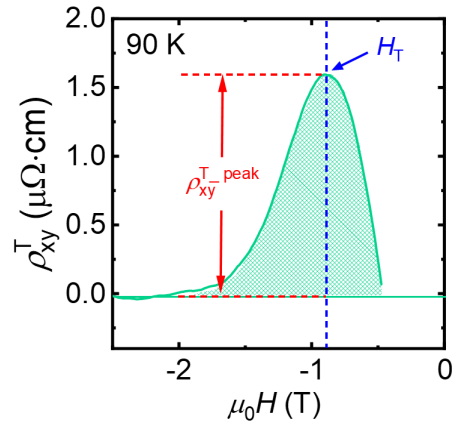

**Figure S6.** The definitions of the $\rho_{xy}^{T\_peak}$ and $H_T$ of the THE. The $\rho_{xy}^{T\_peak}$ represents the amplitude of the THE in $\rho_{xy}$. The $H_T$ marks the position of the maximum value of the THE hump in $\rho_{xy}$.

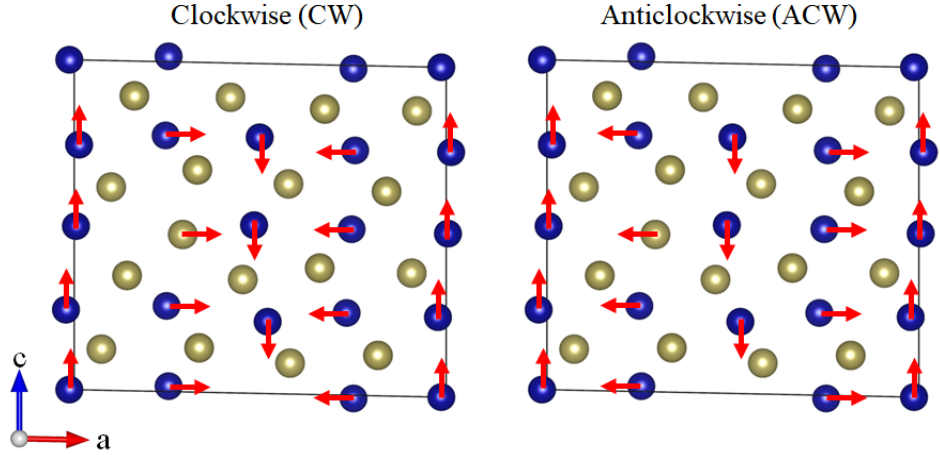

**Figure S7.** The spin textures of the 2 × 1 × 1 supercell of $Cr_5Te_6$ (1.2% vacancy of Cr) with clockwise (CW) and anticlockwise (ACW) chiral spin configurations. Nineteen atoms in the supercell are considered during the DFT first-principles calculations.

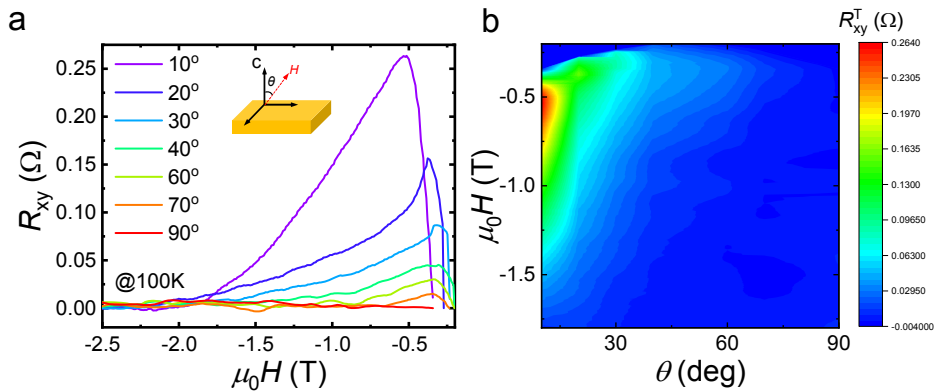

**Figure S8.** a) The magnetic-field dependence of the THE hump at various applied field angles ($\theta$) at 100 K, where $\theta$ is the angle between the magnetic field and $c$ axis. b) The corresponding color map of the data in (a). The hump in the Hall measurements gradually decays as the field is tilted to the $ab$ plane and exists until $\theta$ = 70º. It indicates that the magnetization by the vertical component of the magnetic field is related to the THE in the $Cr_5Te_6$ films,[3] very different from that of the 2D skyrmions.[4]